\documentclass[journal]{IEEEtran}

\ifCLASSINFOpdf
\else
   \usepackage[dvips]{graphicx}
\fi
\usepackage{url}

\usepackage{graphicx}
\usepackage{balance}
\usepackage{amsmath}
\usepackage[pdftex,pdftitle={Very Low Complexity Speech Synthesis Using Framewise Autoregressive GAN (FARGAN) with Pitch Prediction},pdfauthor={Jean-Marc Valin, Ahmed Mustafa, Jan Buethe}]{hyperref}

\begin{document}

\title{Very Low Complexity Speech Synthesis Using Framewise Autoregressive GAN (FARGAN) with Pitch Prediction}

\author{Jean-Marc Valin, \IEEEmembership{Member, IEEE}, Ahmed Mustafa, Jan B\"{u}the \IEEEmembership{Member, IEEE}
\thanks{Jean-Marc Valin is with the Xiph.Org Foundation (e-mail: jmvalin@jmvalin.ca).}
\thanks{Ahmed Mustafa and Jan B\"{u}the are with Amazon Web Services (e-mail: ahdmust@amazon.com, jbuethe@amazon.com).}}

\markboth{Accepted in IEEE Signal Processing Letters}
{Accepted in IEEE Signal Processing Letters}
\maketitle

\begin{abstract}
Neural vocoders are now being used in a wide range of speech processing applications.
In many of those applications, the vocoder can be the most complex component, so
finding lower complexity algorithms can lead to significant practical benefits.
In this work, we propose FARGAN, an autoregressive vocoder that takes advantage of long-term
pitch prediction to synthesize high-quality speech in small subframes, without
the need for teacher-forcing.
Experimental results show that the proposed 600~MFLOPS FARGAN vocoder can achieve
both higher quality and lower complexity than existing low-complexity vocoders.
The quality even matches that of existing higher-complexity vocoders.
\end{abstract}


\IEEEpeerreviewmaketitle

\section{Introduction}
\IEEEPARstart{S}{ince} the publication of the original WaveNet~\cite{van2016wavenet}
and SampleRNN~\cite{mehri2016samplernn} vocoders, neural vocoders have found their
way into a wide range of modern audio processing applications.
These including text-to-speech (TTS) synthesis~\cite{shen2018natural},
low-bitrate speech coding~\cite{kleijn2018wavenet},
super resolution~\cite{liu2022neural}, noise suppression~\cite{maiti2019parametric},
speech codec enhancement~\cite{valin2019resynthesis},
and speed-adjustment~\cite{cohen2022speech}.
That brings neural vocoders among the core building blocks in modern speech processing.
While the original
vocoders had a complexity prohibiting most real-time uses, further improvements
such as WaveRNN~\cite{kalchbrenner2018efficient} and LPCNet~\cite{valin2019lpcnet}
made such applications possible.

The aforementioned autoregressive vocoders all rely on explicit density estimation to
synthesize the speech waveform through conditional sampling, leading to two limitations.
First, their structure requires the use of teacher-forcing~\cite{williams1989learning},
which leads to a \textit{exposure bias}~\cite{schmidt2019generalization}, a domain gap between training and inference that sometimes limits quality.
Second, it prevents direct signal generation and the use of more advanced loss functions,
as done by GAN~\cite{goodfellow2014generative}
vocoders such as MelGAN~\cite{kumar2019melgan}, HiFi-GAN~\cite{kong2020hifi}, and BigVGAN~\cite{lee2022bigvgan}.
On the other hand, according to~\cite{morrison2021chunked},
``autoregressive models possess an inductive bias towards learning pitch and phase". The authors argue
that the phase evolution of a periodic signal is analogous to the cumulative sum problem which is
easier to learn for autoregressive models than for e.g. CNNs which are limited by their finite receptive field.
The authors'
proposed CARGAN model uses past context as implicit pitch conditioning and is 
shown to more accurately represent the pitch compared to the
non-autoregressive HiFi-GAN.
Although CARGAN relies on teacher-forcing with respect to its autoregressive component,
it can still be trained adversarially provided
that the chunk size is sufficiently large.

This paper attempts to further improve the efficiency of autoregressive GAN vocoders
by expanding on both the CARGAN model and our previous Framewise
WaveGAN (FWGAN)~\cite{mustafa2023framewise} vocoder.
We propose (Sec.~\ref{sec:fargan}) a framewise autoregressive GAN (FARGAN) that
explicitly uses pitch-based long-term prediction as a second autoregressive feedback
to improve quality and reduce complexity.
Synthesizing speech based on 2.5~ms subframes to make optimal use of pitch prediction,
we avoid teacher-forcing while still training on sufficiently long sequences by
\textit{unrolling} the model at training time (Sec.~\ref{sec:training}).
The resulting FARGAN model has a size of 820k parameters and a complexity of 600 MFLOPS.
We show in Sec.~\ref{sec:results} that it provides significantly higher quality than the
low-complexity vocoders like LPCNet and Framewise WaveGAN. The quality of FARGAN is even
comparable to that of CARGAN and HiFi-GAN v1 whose complexity is more than 50 times higher.

\section{FARGAN Overview}\label{sec:fargan}

\begin{figure}
\begin{centering}
\includegraphics[width=0.45\columnwidth]{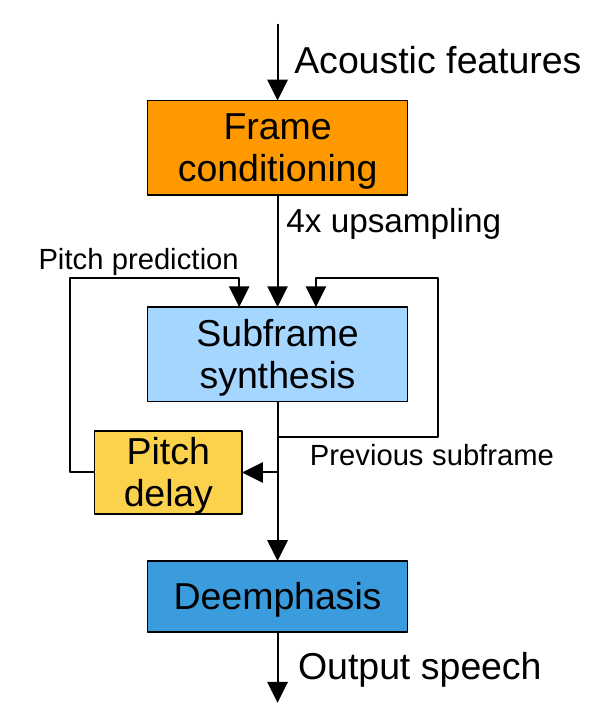}
\par\end{centering}
\caption{Overview of FARGAN. The frame conditioning network operates on acoustic features at a 10-ms interval and outputs a conditioning latent representation at 2.5-ms interval for the autoregressive subframe synthesis network.\label{fig:fargan-overview}}
\end{figure}

Although it can be used in a wide range of applications, FARGAN is designed to
meet the more stringent constraints of real-time speech communications applications.
For those applications, a vocoder needs to produce high-quality speech with a
low algorithmic delay ($< 20~\text{ms}$) and with sufficiently low complexity to avoid
limiting battery life when used continuously on a mobile device CPU.

FARGAN operates on 20-dimensional acoustic features computed at a 10-ms interval on 16~kHz audio.
Each frame is subdivided into 4~subframes of 2.5~ms each, or 40~samples.
As for LPCNet, the acoustic features include 18~Bark-frequency cepstral
coefficients (BFCC), a pitch period, and a voicing indicator.
In each iteration, the model shown in Fig.~\ref{fig:fargan-overview} computes the output for an entire subframe based on
the acoustic features, the previously synthesized subframe, as well as a long term (pitch)
prediction based on the signal history.
The frame conditioning network consists of one fully-connected layer, one 3x1 convolutional layer,
and one transposed convolution layer for performing 4x up-sampling to the subframe rate.
To make the pitch representation easier to learn, the conditioning network also receives
a 12-dimensional embedding of the pitch similar to~\cite{valin2019lpcnet}, for a total input
dimension of 32.

\begin{figure}
\begin{centering}
\includegraphics[width=0.75\columnwidth]{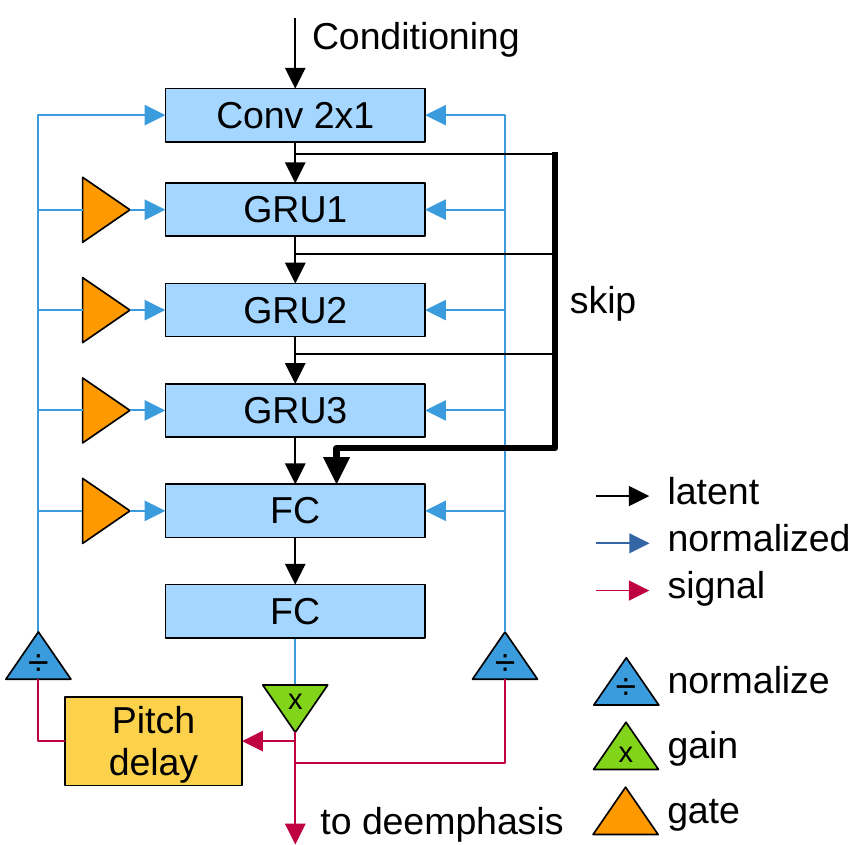}
\par\end{centering}
\caption{Overview of the FARGAN subframe synthesis network.
Multiple inputs to a layer denotes concatenation.
All gains are computed from a small fully-connected layer using the conditioning as input.
The normalization operations apply the inverse of the gain corresponding to the frame where
the signal is used.\label{fig:fargan-subframe}}
\end{figure}

The subframe network shown in Fig.~\ref{fig:fargan-subframe} is responsible for the autoregressive
property of FARGAN.
The signal produced at the subframe network's output feeds back into its input in two ways.
First, the last generated subframe is directly fed back to the input as a way of ensuring
continuity of the signal.
That autoregressive behavior is similar to that of CARGAN~\cite{morrison2021chunked},
except for the much smaller durations involved (40~samples instead of 512).
The second and most important feedback path involves pitch.
The input pitch period is not only used for conditioning, but is also directly used to look back
in the synthesis history and extract the signal exactly one pitch period before the current subframe.
For voiced speech, those samples tend to be an accurate prediction for the current subframe
being synthesized.
However, as a consequence of using pitch prediction directly, the proposed model cannot
be easily adapted to synthesize general audio and music.

To handle cases where the pitch period $T$ is shorter than the subframe size $N$, 
the predicted signal $p(n)$ is obtained by
\begin{equation}
p(n) = \begin{cases}
\hat{x}(n-T) & T \ge N\\
\hat{x}(n-2T) &\text{otherwise}.
\end{cases}
\end{equation}
The period $T$ can never be shorter than $N/2$ since the highest pitch allowed is 500~Hz, corresponding to $T=32$.

All layers of the subframe network use a $\tanh()$ activation. With the exception of the output layer,
all layers of the subframe network also include a gated linear unit (GLU)
\begin{equation}
G(\mathbf{x}) = \mathbf{x}\odot \sigma{\left(\mathbf{Wx}\right)}
\end{equation}
at their output, where $\odot$ denotes the Hadamard product, $\sigma(\cdot)$
denotes the sigmoid function, and $\mathbf{W}$ is the GLU weight matrix.

We use normalization to reduce the dynamic range of the synthesized signal.
For each subframe, a single fully-connected neuron with exponential activation computes
a gain from the subframe conditioning input.
That gain is applied to the subframe output layer to scale the
synthesized speech to its full dynamic range.
In the autoregressive feedback, the previous subframe and pitch prediction are
renormalized using the gain of the subframe where they are used rather than the one
where the speech was generated.

We find that simply feeding the autoregressive components back to the input
of the subframe network does not result in optimal use of that information --
likely due to vanishing gradient.
For that reason we also feed them to all the other subframe layers.
Similar to skip connections, we find that although it does not significantly
improve the final output quality, it helps stabilize and speed up convergence.
In the case of the pitch prediction, we also add a gate that avoids the
prediction being used for unvoiced speech.
The gate values are computed from the conditioning in the same way as the
gain described above.

The output of the subframe network is de-emphasized with the first-order IIR filter
$H(z)=1/(1-\alpha z^{-1})$, with $\alpha=0.85$. Operating in the pre-emphasized
domain reduces the precision required in the internal representation and,
combined with the gain normalization described above, makes it possible
to use 8-bit quantization throughout the model for both the weights and the activations.

\subsection{Computational Considerations}

To make FARGAN useful on a wide range of devices, we need to reduce both the number
of operations (multiplications and additions) required, but also the size of the model.
Fewer operations has obvious benefits in terms of speed and power consumption,
but model size is also very important.
A smaller model will not only reduce the cache/memory bandwidth required,
but also allow the model to be held in a smaller and faster cache.
Moreover, reducing the size of each weight -- in our case from 32-bit floating-point
to 8-bit integers -- makes it possible to compute 4~times more operations at a time
for the same SIMD (single instruction, multiple data) vector length.

The choice of $\tanh()$ and sigmoid activation above is motivated by the fact that their
$\pm1$ bounds make it easy to quantize to 8~bits (unlike the unbounded ReLU).
Similarly, the use of 2.5~ms subframes compared to the 10-ms subframes in our previous
Framewise WaveGAN work further reduces the model size for a given complexity.
The proposed model weights can thus fit in less than
1~MB, corresponding to the L2 cache of newer CPUs, or the L3 cache of older CPUs.

\section{Training}\label{sec:training}

Unlike many other autoregressive vocoders, FARGAN training does not (and cannot
due to its structure) involve teacher forcing~\cite{williams1989learning}.
Instead, the subframe network is \textit{unrolled} in time in such a way that the autoregressive
components used at training time are based on the synthesized speech, rather than the ground truth speech.
The authors are aware of several unsuccessful attempts (including their own) at adding direct pitch
prediction to enhance the efficiency of LPCNet.
One of the likely culprits for those failures is the use of teacher forcing.
That is another reason for seeking to avoid teacher forcing in FARGAN.
Due to the small model size and the framewise generation, training the unrolled model
is still fast enough.

\subsection{Pretraining}

Let $X_L(\ell, k)$ denote the short-time Fourier transform (STFT) of signal $x$ with
window size $L$ for frame $\ell$ at frequency $k$ and 75\% overlap.
We define the spectral loss $\mathcal{L}_L$ between the ground truth signal $x$ and
the synthesized signal $\hat{x}$ as
\begin{equation}
\mathcal{L}_L = \sum_\ell \sum_k \left| |\hat{X}_L(\ell, k)|^\gamma - |X_L(\ell, k)|^\gamma \right|\,,
\end{equation}
where $\gamma=0.5$ approximates the perceived loudness~\cite{moore2012introduction}.

For pre-training, we use a multi-resolution spectral loss
\begin{equation}
\mathcal{L}^{(S)} = \mathcal{L}_{80} + \mathcal{L}_{160} + \mathcal{L}_{320} + \mathcal{L}_{640} + \mathcal{L}_{1280} + \mathcal{L}_{2560}\,.
\label{eq:spectral-loss}
\end{equation}

In the pretraining phase, we use sequences of 15~frames, with 10\% of the sequences being 30-frame long.
Pre-training for 470k~updates with sequences of 15~frames and a batch size of 4096
takes approximately 2.5~days on one Nvidia A100 GPU.

\subsection{Adversarial Training}
For adversarial training, we use multi-resolution magnitude-STFT discriminators similar to \cite{jang2021univnet}. This choice is motivated by the observation
that otherwise popular time domain-discriminators (\cite{kumar2019melgan, kong2020hifi, 9413605}) for vocoder training failed
to improve (and in fact even degraded) quality for two previous block-wise signal processing models, namely the FWGAN vocoder \cite{mustafa2023framewise}
and the NoLACE enhancement model \cite{buethe2024nolace}. In both cases, the multi-scale and the multi-period discriminators from \cite{kong2020hifi}
would quickly win against the generators indicating that these discriminators are capable of detecting (potentially irrelevant) irregularities in the
generated signals that the generators were not able to remove to a satisfactory degree. A possible explanation for this behavior could be the block-wise
signal processing itself and the fact that small temporal irregularities are easier detected from a raw time-domain signal than from a log-magnitude spectrogram.

We follow the architecture from~\cite{mustafa2023framewise} and use 6~STFT
discriminators $D_k$ with the modifications described in \cite{buethe2024nolace}:
Each discriminator takes as input a log-magnitude spectrogram
computed from size-$2^{k+5}$ STFTs with 75\% overlap.
To simplify notation, we use $D_k(x)$ and $D_k(\hat x)$, treating the
log-magnitude STFT transform of $x$ and $\hat{x}$ as part of the discriminator.
We furthermore apply strides along the frequency axis to keep the frequency
range of the receptive fields constant.
This has been found to increase the ability of discriminators with high
frequency resolution to detect inter-harmonic noise.
Finally, we concatenate a two-dimensional frequency positional sine-cosine
embedding to the input channels of every 2d-convolutional layer.

We train FARGAN as a least-squares GAN~\cite{mao2017least}.
First we note that the generated signal $\hat x$ depends deterministically on ground-truth signal $x$. With this we define the adversarial part of the training loss for FARGAN as
\begin{equation}
    \mathcal{L}_{\textrm{adv}}(x, \hat x) = \frac{1}{6}\sum_{k=1}^6 E_{x}[(1 - D_k(\hat x))^2] + \mathcal{L}_{\mathrm{feat}}(D_k, x, \hat x),
\end{equation}
where $\mathcal{L}_{\mathrm{feat}}$ denotes the standard feature matching loss, i.e. the mean of the $L_1$ losses of hidden layer outputs for $x$ and $\hat x$.

The complete training loss for FARGAN includes the pre-training spectral loss, such that
\begin{equation}
    \mathcal{L}_{\textrm{tot}}(x, \hat x) = \mathcal{L}_{\textrm{adv}}(x, \hat x) + \mathcal{L}^{(S)}(x, \hat x).
\end{equation}

Simultaneously, the discriminators are trained to minimize the loss
\begin{equation}
    \mathcal{L}_{D_k}(x, \hat x) = E_{x}[D_k(\hat x)^2 + (1 - D_k(x))^2].
\end{equation}

Adversarial training is carried out on 60-frame sequences for another 50~epochs with a fixed learning rate of~$2\cdot 10^{-6}$ and a batch size of~160, which corresponds to about 380k training steps. We used the Adam optimizer with $\beta_1=0.9$ and $\beta_2=0.999$ for both FARGAN and the discriminators.

\section{Experiments \& Results}\label{sec:results}
We train speaker-independent FARGAN models on 205~hours of 16-kHz speech from a combination of TTS
datasets~\cite{demirsahin-etal-2020-open,kjartansson-etal-2020-open,kjartansson-etal-tts-sltu2018,guevara-rukoz-etal-2020-crowdsourcing,he-etal-2020-open,oo-etal-2020-burmese,van-niekerk-etal-2017,gutkin-et-al-yoruba2020,bakhturina2021hi}
including more than 900~speakers in 34~languages and dialects.

We evaluate two versions of FARGAN: a proposed 820k-weight model and an even smaller 500k-weight model.
As an ablation study, we evaluate the effect of removing (from the larger proposed model)
the pitch prediction (replacing it with a larger history to maintain the same number of weights)
and removing all autoregressive behavior.

We compare FARGAN to three other low-complexity vocoders: LPCNet~\cite{valin2019lpcnet}, Framewise
WaveGAN~\cite{mustafa2023framewise}, and HiFi-GAN~\cite{kong2020hifi}~v3.
As a reference, we also include CARGAN and HiFi-GAN~v1, which have a much higher-complexity
than the proposed vocoder.
All evaluations are conducted at 16~kHz and all vocoders are trained using the same datasets
as FARGAN.
The complete implementation of FARGAN is available under an open-source
license\footnote{\href{https://gitlab.xiph.org/xiph/opus/-/tree/spl\_fargan/dnn/torch/fargan}{https://gitlab.xiph.org/xiph/opus/-/tree/spl\_fargan/dnn/torch/fargan}}.

\begin{table}
    \caption{Objective evaluation of the different vocoders using PESQ, WARP-Q and mean pitch error (MPE)}
    \label{tab:objective}
    \centering
     \begin{tabular}{llll}
    \hline
    Condition & PESQ    & WARP-Q & MPE \\ \hline
\textbf{FARGAN} &  \textbf{3.298} & 0.587 & \textbf{4.108} \\
\hspace{2em}small &  3.241 & 0.615 & 4.172 \\
\hspace{2em}no-pitch& 3.174 & 0.608 & 4.239 \\
\hspace{2em}no-AR  &  2.859 & 0.655 & 4.457 \\
CARGAN       &  3.127 & 0.559 & 4.322 \\
HiFi-GAN v1   &  3.024 & \textbf{0.495} & 5.501 \\
HiFi-GAN v3   &  2.373 & 0.651 & 6.715 \\
LPCNet       &  2.539 & 0.694 & 5.303 \\
FWGAN        &  2.833 & 0.648 & 5.063 \\ \hline
    \end{tabular}
\end{table}

We evaluated the algorithms using 192~clean English speech clips from
the NTT Multi-Lingual Speech Database for Telephonometry and 192~clean English clips from
the PTDB-TUG~\cite{ptdb} database.
No items from these databases were included in the training data.
Audio samples synthesizing clean speech, singing voice and noisy speech samples are available on a demo page
\footnote{\href{https://ahmed-fau.github.io/fargan\_demo/}{https://ahmed-fau.github.io/fargan\_demo/}}.
These demonstrate that FARGAN generalizes well to unseen conditions considering that no singing or noisy speech
was used at training time.

We first evaluated all the vocoders objectively using PESQ~\cite{P.862}, WARP-Q~\cite{jassim2021warp},
as well as mean pitch accuracy (MPE), as measured in~\cite{mustafa2023framewise}.
Table~\ref{tab:objective} shows the objective evaluation results,
demonstrating that FARGAN achieves better pitch accuracy than all other vocoders.
All three objective metrics are in agreement in demonstrating the effectiveness of the autoregressive components and show
that explicit pitch prediction can help achieve both higher quality and better pitch accuracy.
Although objective metrics are designed to correlate with subjective quality to a certain degree,
comparing very different families of algorithms is a notoriously difficult task for these metrics.
In this case, we notice that PESQ and WARP-Q give opposite rankings for FARGAN, HiFiGAN v1,
and CARGAN, so a subjective evaluation is needed.

\begin{figure}
\begin{centering}
\includegraphics[width=0.8\columnwidth]{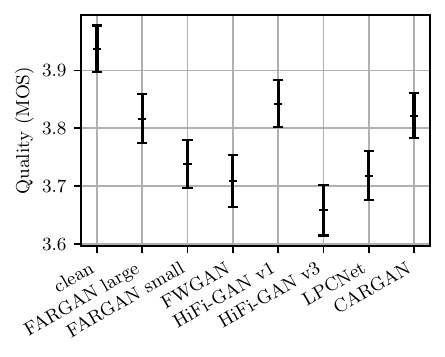}
\par\end{centering}
\caption{P.808 mean opinion score (MOS) results including the 95\% confidence intervals.
FARGAN large, HiFi-GAN v1 and CARGAN are statistically tied and out-perform all other vocoders
with $p<0.05$. \label{fig:mos-results}}
\end{figure}

For subjective evaluation, we used the crowd-sourcing methodolgy from ITU-R P.808\cite{P.808}.
Each sample was subjectively evaluated by 9~randomly-selected naive listeners.
Results in Fig.~\ref{fig:mos-results} show that the larger FARGAN model
is statistically tied with CARGAN and HiFi-GAN~v1, and significantly better ($p<0.05$) than
LPCNet, FWGAN, and HiFi-GAN~v3.
Moreover, the smaller FARGAN model is statistically tied with LPCNet and FWGAN,
and out-performs ($p<0.05$) HiFi-GAN~v3.

\subsection{Complexity}

\begin{table}
    \caption{Complexity of the different vocoders. The number of operations, expressed
    in GFLOPS counts one multiply-add operation as two FLOPS.
    When available, we also include the percentage of one i7-8565 CPU core required
    for real-time operation (inverse of "real-time factor").}
    \label{tab:complexity}
    \centering
     \begin{tabular}{lll}
    \hline
    Condition & GFLOPS  & CPU (\%)  \\ \hline
\textbf{FARGAN} &  0.6 & 0.8  \\
\hspace{2em}small &  0.35 & 0.5  \\
CARGAN       &  65.9 & -  \\
HiFi-GAN v1   &  38.1 & -  \\
HiFi-GAN v3   &  2.8 & -  \\
LPCNet       &  2.8 & 4.5  \\
FWGAN        &  1.2 & -  \\ \hline
    \end{tabular}
\end{table}

Table~\ref{tab:complexity} compares the complexity of the different vocoders, both
in number of operations and speed on actual hardware (when available).
The proposed FARGAN model has a complexity of 0.6~GFLOPS, which about 5~times less
complex than LPCNet and HiFi-GAN~v3, despite providing a higher quality.
Compared to the high-quality CARGAN and HiFi-GAN~v1, FARGAN
achieves a complexity reduction of 110x and 64x repectively, with equivalent quality.
Using an optimized C implementation, FARGAN can
synthesize speech in real time using less than 1\% of a modern laptop or phone CPU core.

\section{Conclusion}

We have demonstrated a very-low-complexity GAN vocoder that uses pitch-prediction-based autoregression to achieve
high-quality speech synthesis using only 600~MFLOPS.
The proposed FARGAN vocoder achieves both higher quality and lower complexity
when compared to other existing low-complexity vocoders (LPCNet, HiFi-GAN v3, FWGAN).
Moreover, it matches the quality of state-of-the-art high-complexity vocoders (HiFi-GAN v1, CARGAN).
We believe the demonstrated reduction in vocoder complexity opens the way for neural vocoders
being used in new applications, including low-power embedded systems.

\balance

\bibliographystyle{IEEEbib}
\bibliography{corpora,lpcnet}

\end{document}